\documentclass[aps,pre,twocolumn,floats,amsmath,amssymb,showpacs,floatfix]%
{revtex4}
\usepackage{graphicx}
\usepackage{dcolumn}

\begin{document}
\title{Correlation Dimension of Inertial Particles in Random Flows}
\author{M. Wilkinson$^{(1)}$, B. Mehlig$^{(2)}$ and K. Gustavsson$^{(2)}$}
\affiliation{$^{(1)}$Department of Mathematics and Statistics,
The Open University, Walton Hall, Milton Keynes, MK7 6AA, England\\
$^{(2)}$Department of Physics, Gothenburg University, 41296
Gothenburg, Sweden}

\begin{abstract}
We obtain an implicit equation for the correlation dimension $D_2$ of dynamical systems in terms
of an integral over a propagator. We illustrate the utility of this approach by evaluating $D_2$ for inertial particles suspended in a random flow. In the limit where the correlation time of the flow field approaches zero, taking the short-time limit of the propagator enables $D_2$ to be determined from the solution of a partial differential equation. We develop the solution as a power series in a dimensionless parameter which represents the strength of inertial effects.
\end{abstract}
\pacs{05.40.-a,05.45-a}

\maketitle

The behaviour of small particles moving independently in complex flows is a fundamental problem in fluid mechanics, which has applications in understanding rainfall \cite{Sha03}, planet formation \cite{Beck+00,Wil+08} and many areas of technology and environmental science. It is known that when the inertia of the particles is significant, clustering may occur \cite{Max87}, which can lead to an increase in the rate of collision or aggregation of the particles, and which can also affect the scattering of electromagnetic radiation. In developing a description of these processes the most natural way to quantify the clustering is to consider the number of particles ${\cal N}$ inside a ball of radius $\delta r$ centred on any given particle. If this quantity has a power-law dependence for small $\delta r$ of the form ${\cal N}\sim \delta r^{D_2}$ (with $D_2$ less than the dimension of space, $d$), the particles cluster onto a fractal attractor. The quantity $D_2$ is termed  the correlation dimension \cite{Ott02}. The clustering process is in fact found to approach a fractal attractor \cite{Som+93}.

It is desirable to develop a theoretical understanding of the clustering effect. It has been ascribed to particles (which we assume to be much denser than the fluid) being centrifuged away from vortices \cite{Max87}, but other explanations (for example, caustics \cite{Fal+02,Wil+05}) are possible. In particular, a model with a short-time correlated velocity field, analysed in \cite{Wil+07},  gives good agreement with a numerical determination of the Lyapunov dimension $D_{\rm L}$ of particles in Navier-Stokes turbulent flow, reported in \cite{Bec+06} (the Lyapunov dimension was introduced in \cite{Kap+79}, and is discussed in \cite{Ott02}). The task of calculating the more physically interesting dimension $D_2$ by analytical methods has appeared to be intractable, but we show that $D_2$ can be obtained more easily than $D_{\rm L}$. We give a general prescription for calculating the correlation dimension, which can also be applied to other types of dynamical system. We show that when the turbulent velocity is modelled by a random vector field with a short correlation time (that is, for the model analysed in \cite{Wil+07}), this leads to an expansion of $D_2$ as a power series in a parameter $\epsilon$ which is a dimensionless measure of the inertia of the particles. The coefficients of this series may be obtained exactly to arbitrarily high order. We show how convergent results are obtained using a conformal Borel summation.

The correlation dimension $D_2$ may be defined in terms of the expected number ${\cal N}(\delta r)$ of particles inside a ball of radius $\delta r$ surrounding a test particle:
\begin{equation}
\label{eq: 1}
D_2=\lim_{\delta r \to 0}\frac{{\rm ln}[\langle {\cal N}(\delta r)\rangle]}{{\rm ln}(\delta r)}
\end{equation}
(where $\langle X\rangle$ denotes an average of $X$), provided this limit exists and satisfies $D_2\le d$, where $d$ is the dimensionality of space. This implies that $\langle {\cal N}(\delta r)\rangle \sim \delta r^{D_2}$ which is the radial part of the volume element of a ball in $D_2$ dimensions. If the limit in (\ref{eq: 1}) is greater than or equal to $d$, there is no clustering, and $D_2=d$. While $D_2$ has fundamental importance, it is difficult to calculate analytically. It can be expressed in terms of the large deviation statistics of the finite-time Lyapunov exponents, $\sigma(t)$ \cite{Gra+84,Ott02,Bec+04}. These statistics are very difficult to calculate by means other than numerical simulations (although they have been evaluated for the Kraichnan model for advection in short-time correlated flows \cite{Bec+04}). Earlier studies of $D_2$ for particles with significant inertia have been numerical evaluations \cite{Bec+07,Bec+07a}.

We consider the motion of small, dense particles suspended in a turbulent fluid with velocity field $\mbox{\boldmath$u$}(\mbox{\boldmath$r$},t)$. The motion of a particle at position $\mbox{\boldmath$r$}$ moving with velocity $\mbox{\boldmath$v$}$ is determined by viscous damping of the particle relative to the fluid. The equations of motion are
\begin{equation}
\label{eq: 2}
\dot{\mbox{\boldmath$r$}}=\mbox{\boldmath$v$}
\ ,\ \ \
\dot{\mbox{\boldmath$v$}}=-\gamma [\mbox{\boldmath$v$}-\mbox{\boldmath$u$}(\mbox{\boldmath$r$}(t),t)]
\end{equation}
where we use the notation $\dot X={\rm d}X/{\rm d}t$ and where $\gamma$ is a damping rate proportional to the viscosity. In this paper we consider how to extract information about $D_2$ from a quantity $Z_1(t)$ which is defined to be the logarithmic derivative of the separation $\delta r$ between two particles:
\begin{equation}
\label{eq: 3}
\frac{\delta \dot r}{\delta r}=Z_1(t)
\ .
\end{equation}
An equation of motion for $Z_1$ which is valid when $\delta r$ is sufficiently small may be obtained from the linearisation of (\ref{eq: 2}) as discussed below: $Z_1(t)$ may be coupled to one or more additional variables $Z_2(t),\ldots$, but the equations for the $Z_i$ are independent of $\delta r$ provided that quantity is sufficiently small. We also consider the variable
\begin{equation}
\label{eq: 4}
Y(t)={\rm ln}\,\delta r(t)
\end{equation}
which is related to $Z_1$ by $\dot Y=Z_1$. Note that $Y$ is related to the finite-time Lyapunov exponent $\sigma(t)$ at time $t$: we have $Y(t)-Y(0)=t\sigma(t)$ (provided $\delta r$ is everywhere sufficiently small). In the following we discuss the two-dimensional case where $Z_1$ is coupled to one additional variable $Z_2$. We consider the joint probability density $\rho(Y,Z_1,Z_2)$ of $Y$, $Z_1$ and $Z_2$. Because the equation of motion of $Z_1$ and $Z_2$ is independent of $Y={\rm ln}\,\delta r$ when the linearised equation is valid, in the steady state the joint distribution factorises, with the distribution of $Y$ being in a form which reflects the translational invariance in $Y$. Because the eigenfunctions of translations are exponential functions, the steady-state joint distribution of $Y$, $Z_1$, $Z_2$ is
\begin{equation}
\label{eq: 5}
\rho(Y,Z_1,Z_2)=\exp(\alpha Y)\rho_Z(Z_1,Z_2)
\end{equation}
for some constant $\alpha$. This form is not normalisable, but it should be remembered that (\ref{eq: 5}) is only valid when $\delta r$ is sufficiently small. In the case where $\alpha>0$, the form (\ref{eq: 5}) can be matched to a distribution which is valid for large $\delta r$ to make a normalisable solution, whereas $\alpha<0$ is not allowed.
The distribution (\ref{eq: 5}) implies that the distribution of $Y$ has probability element ${\rm d}P=\exp(\alpha Y){\rm d}Y=\delta r^{\alpha-1}{\rm d}\delta r$. The relation (\ref{eq: 1}) implies that the probability for the separation to be in an interval ${\rm d}\delta r$ is ${\rm d}P=\delta r^{D_2-1}{\rm d}\delta r$, so that
\begin{equation}
\label{eq: 6}
D_2=\alpha
\ .
\end{equation}

The condition for determining $D_2=\alpha$ is that this distribution (\ref{eq: 5}) should be invariant under time evolution. This is expressed in terms of a propagator for the time-evolution of $Y$ and $\mbox{\boldmath$Z$}=(Z_1,Z_2)$. Specifically, this propagator $K(\Delta Y,\mbox{\boldmath$Z$},\mbox{\boldmath$Z$}',\Delta t)$ is defined to be the probability density for $Y$ to change by $\Delta Y$ and for $\mbox{\boldmath$Z$}=(Z_1,Z_2)$ to change from $\mbox{\boldmath$Z$}'$ to $\mbox{\boldmath$Z$}$ in time $\Delta t$. Stationarity of the distribution (\ref{eq: 5}) then leads to
\begin{equation}
\label{eq: 7}
\rho_Z(Z_1,Z_2)=\int_{-\infty}^\infty {\rm d}\Delta Y \int_{-\infty}^\infty {\rm d}Z_1' \int_{-\infty}^\infty {\rm d}Z_2'
\nonumber
\end{equation}
\begin{equation}
\times \exp(-\alpha \Delta Y)\,K(\Delta Y,\mbox{\boldmath$Z$},\mbox{\boldmath$Z$}',\Delta t)\,\rho_Z(Z_1',Z_2')
\end{equation}
which is satisfied for all $\Delta t$. In the case $\Delta t\to \infty$, the propagator $K$ is related to the large-deviation probability density function for the finite-time Lyapunov exponent. This leads to a formulation (to be discussed in a later paper) which is equivalent to some earlier theories for determining $D_2$ \cite{Gra+84,Ott02,Bec+04}. Here, however, we concentrate upon the short-time limit, $\Delta t\to 0$. We shall see that this leads to an analysis of $D_2$ in terms of a differential equation, which is much more analytically tractable.

To make further progress we need to consider the equation of motion for the variables $Z_1,Z_2$ in the two-dimensional case. Parts of the calculation follow \cite{Meh+04}, but here we use a simpler operator algebra. The linearised equations of motion corresponding to (\ref{eq: 2}) are $\delta \dot {\mbox{\boldmath$r$}}=\delta {\mbox{\boldmath$v$}}$ and $\delta \dot {\mbox{\boldmath$v$}}=-\gamma \delta {\mbox{\boldmath$v$}}+\gamma{\bf E}(t)\delta
{\mbox{\boldmath$r$}}$ where ${\bf E}(t)$ is a $2\times 2$ matrix with elements
$E_{ij}(t)={\partial u_i/{\partial r_j}}({\mbox{\boldmath$r$}}(t),t)$.
We write $\delta {\mbox{\boldmath$r$}}=\delta r {\bf n}_\theta$ and
$\delta {\mbox{\boldmath$v$}}=Z_1 \delta r {\bf n}_\theta+Z_2 \delta r {\bf n}_{\theta+\pi/2}$,
where ${\bf n}_\theta$ is unit vector in direction $\theta$. Expressing the linearised equations of motion in terms of the variables $\delta r$, $Z_1$, $Z_2$ we obtain \cite{Meh+04}
\begin{eqnarray}
\label{eq: 8}
\dot Z_1&=&-\gamma Z_1+(Z_2^2-Z_1^2)+\gamma E_{\rm d}(t)
\nonumber \\
\dot Z_2&=&-\gamma Z_2-2Z_1Z_2+\gamma E_{\rm o}(t)
\end{eqnarray}
where $E_{\rm d}(t)={\bf n}_{\theta}\cdot{\bf E}(t){\bf n}_{\theta}$ and $E_{\rm o}(t)={\bf n}_{\theta+\pi/2}\cdot{\bf E}(t){\bf n}_\theta$, and $\delta \dot r=Z_1 \delta r$, $\dot \theta=Z_2$ (so that the definition of $Z_1$ is consistent with (\ref{eq: 3})). It might be expected that the distribution of $(Z_1,Z_2)$ obtained from the long-time limit of the evolution of equation (\ref{eq: 8}), which we term $\rho_0(Z_1,Z_2)$, is the same as the distribution $\rho_Z(Z_1,Z_2)$ in (\ref{eq: 7}). However, $\rho_Z$ differs from $\rho_0$ because it is conditioned upon being at a particular value of $Y$. If $\alpha> 0$, particles reaching a negative value of $Z_1$ arrive from a larger value of $Y$, where the probability, density is larger. This implies that the distributions $\rho_0$ and $\rho_Z$ are different, and that $\rho_Z$ has a smaller mean value of $Z_1$ than $\rho_0$.

Next we must specify a model for the two-dimensional velocity field $\mbox{\boldmath$u$}(\mbox{\boldmath$r$},t)$. We allow this to be partially compressible by writing $\mbox{\boldmath$u$}=\mbox{\boldmath$\nabla$}\Phi+\mbox{\boldmath$\nabla$}\wedge \Psi{\bf e}_3$. In order to use statistical techniques we consider the stream function $\Psi(\mbox{\boldmath$r$},t)$ and potential $\Phi(\mbox{\boldmath$r$},t)$ to be random scalar fields with specified correlation functions. We shall assume that $\langle\Phi(\mbox{\boldmath$r$},t)\Phi(\mbox{\boldmath$r$}',t')\rangle=C(\vert\mbox{\boldmath$r$}-\mbox{\boldmath$r$}'\vert,\vert t-t'\vert)$, where $C(R,t)$ has support $\xi$ (the correlation length) and $\tau$ (the correlation time) in $R$ and $t$ respectively. Also, we assume that $\Phi$ and $\Psi$ are uncorrelated and that the correlation function of $\Psi$ is proportional to that of $\Phi$, such that $\langle \Psi^2\rangle/\langle \Phi^2\rangle=\beta^2$ for some number $\beta$. Furthermore, in this paper we consider the limit where we the correlation time $\tau$ is sufficiently small that the randomly fluctuating terms in (\ref{eq: 8}), $E_{\rm d}(t)$ and $E_{\rm o}(t)$, can be treated as white noise. In this case the equations of motion for $Z_1$, $Z_2$ become a pair of coupled Langevin equations, and the probability density $\rho_0(Z_1,Z_2)$ generated by equation (\ref{eq: 8}) obeys a diffusion equation, which can be written formally as
\begin{equation}
\label{eq: 9}
\frac{\partial \rho_0}{\partial t}=\hat {\cal F}_0 \rho_0
\end{equation}
where $\hat {\cal F}_0$ is a Fokker-Planck operator:
\begin{eqnarray}
\label{eq: 10}
\hat {\cal F}_0 \rho_0 &=&\frac{\partial}{\partial Z_1}[(\gamma Z_1+Z_1^2-Z_2^2)\rho_0]
+{\cal D}_{11}\frac{\partial^2 \rho_0}{\partial Z_1^2}
\nonumber \\
&+&\!\!\!\frac{\partial}{\partial Z_2}[(\gamma Z_2+2Z_1Z_2)\rho_0]
+{\cal D}_{22}\frac{\partial^2 \rho_0}{\partial Z_2^2}
\ .
\end{eqnarray}
Here the diffusion coefficients are expressed in terms of correlation functions of the velocity gradients:
\begin{equation}
\label{eq: 11}
{\cal D}_{ii}=\tfrac{1}{2}\gamma^2\int_{-\infty}^\infty {\rm d}t\ \langle E_{i1}(t)E_{i1}(0)\rangle
\ .
\end{equation}

Now we consider how equations (\ref{eq: 9}), (\ref{eq: 10}) are used to construct the short-time propagator in (\ref{eq: 7}). For small $\Delta t$, $Y$ evolves ballistically, with velocity $Z_1\sim Z_1'$. In the short time limit, the action of the propagator $K(\Delta Y,\mbox{\boldmath$Z$},\mbox{\boldmath$Z$}',\Delta t)$ in (\ref{eq: 8}) on a function $f(Y,Z_1,Z_2)$ can therefore be written as $f_K(Y,Z_1,Z_2)=f(Y-Z_1\Delta t,Z_1,Z_2)+\Delta t\,\hat {\cal F}_0\,f(Y,Z_1,Z_2)+O(\Delta t^2)$. The equation (\ref{eq: 7}) determining self-reproduction of $\rho_Z(Z_1,Z_2)$ therefore becomes $\rho_Z(Z_1,Z_2)=\exp(-\alpha Z_1\Delta t)\rho_Z(Z_1,Z_2)+\Delta t\, \hat {\cal F}_0\,\rho_Z(Z_1,Z_2)+O(\Delta t^2)$. Extracting the $O(\Delta t)$ term gives the differential equation
\begin{equation}
\label{eq: 12}
\alpha Z_1\rho_Z(Z_1,Z_2)-\hat {\cal F}_0\rho_Z(Z_1,Z_2)=0
\ .
\end{equation}
Upon integrating over space, and using the fact that the operator $\hat {\cal F}_0$ is a divergence, we have
\begin{equation}
\label{eq: 13}
\int_{-\infty}^\infty {\rm d}Z_1 \int_{-\infty}^\infty {\rm d}Z_2\ Z_1\,\rho_Z(Z_1,Z_2)=\langle Z_1\rangle=0
\ .
\end{equation}
The value of $D_2$ is determined by finding the value of $\alpha$ for which a normalisable solution of (\ref{eq: 12}) can be obtained for which the mean value of $Z_1$ is zero. The equations (\ref{eq: 12}) and (\ref{eq: 13}) together constitute an exact method for determining $D_2=\alpha$. Their extension to the three-dimensional case is straightforward.

It is useful to make a change of variable from $(Z_1,Z_2)$ to scaled variables $(x_1,x_2)$ defined  by $x_i=\sqrt{\gamma/{\cal D}_{ii}}Z_i$, and to use a dimensionless time $t'=\gamma t$. We also introduce two dimensionless parameters, $\epsilon$, which measures the importance of inertial effects, and $\Gamma$, which is a convenient measure of the relative magnitudes of $\Psi$ and $\Phi$:
\begin{equation}
\label{eq: 14}
\epsilon =\sqrt{\frac{{\cal D}_{11}}{\gamma}}
\ ,\ \ \
\Gamma=\frac{{\cal D}_{22}}{{\cal D}_{11}}=\frac{1+3\beta^2}{3+\beta^2}
\ .
\end{equation}
Using these new variables (\ref{eq: 12}) becomes an equation for the joint probability density $P(x_1,x_2)$ of $x_1$, $x_2$:
\begin{equation}
\label{eq: 15}
\hat F\,P=0=\frac{\partial}{\partial x_1}[(x_1+\epsilon(x_1^2-\Gamma x_2^2))P]
\nonumber
\end{equation}
\begin{equation}
+\frac{\partial}{\partial x_2}[(x_2+2\epsilon x_1x_2)P]+\frac{\partial^2 P}{\partial x_1^2}+\frac{\partial^2 P}{\partial x_2^2}-\epsilon \alpha x_1 P
\end{equation}
(which defines the differential operator $\hat F(\epsilon,\alpha,\Gamma)$).
Equation (\ref{eq: 15}) is to be solved with the supplementary condition $\langle x_1\rangle=0$, which can only be satisfied for isolated values of $\alpha$. Our solution below obtains one unique value of $\alpha$, which is $D_2$.

We now develop the solution as a series expansion in $\epsilon$, using a system of annihilation and creation operators which are analogous to those used in quantum mechanics. We use a notation similar to the Dirac notation, whereby a function $f(x_1,x_2)$ is denoted by a vector $|f)$. We expand both the solution $|P)$ of (\ref{eq: 15}) and the value of $\alpha$ for which the solution of this equation exists and satisfies $\langle x_1 \rangle=0$ as power series in $\epsilon$:
\begin{equation}
\label{eq: 16}
|P ) =\sum_{k=0}^\infty \epsilon^k\, |P_k)
\ ,\ \ \
\alpha = \sum_{k=0}^\infty \epsilon^k\, \alpha_k
\ .
\end{equation}
We write the Fokker-Planck operator in (\ref{eq: 15}) as
\begin{equation}
\label{eq: 17}
\hat F=\hat F_0+\epsilon (\hat G -\alpha \hat x_1)
\end{equation}
(thereby defining operators $\hat F_0$, $\hat G$). The unperturbed steady-state $|P_0)$ satisfying $\hat F_0|P_0)=0$ is $P_0(x_1,x_2)=\exp[-(x_1^2+x_2^2)/2]/2\pi$, and other eigenfunctions of $\hat F_0$ are generated by creation operators $\hat a_i$ and annihilation operators $\hat b_i$:
\begin{equation}
\label{eq: 18}
\hat a_i =-\partial_{x_i}\ ,\ \ \ \hat b_i=\partial_{x_i}+x_i
\ .
\end{equation}
These operators generate eigenfunctions satisfying $\hat F_0|\phi_{nm})=-(n+m)|\phi_{nm})$, according to the rules
\begin{eqnarray}
\label{eq: 19}
\hat a_1 |\phi_{n,m})=|\phi_{n+1,m})&\! &\hat b_1 |\phi_{n,m})=n|\phi_{n-1,m})
\nonumber \\
\hat a_2 |\phi_{n,m})=|\phi_{n,m+1})&\! &\hat b_2 |\phi_{n,m})=m|\phi_{n,m-1})
\end{eqnarray}
with $|\phi_{00})=|P_0)$, which is normalised as a probability density. The states $|P_k)$ in (\ref{eq: 16}) with be expressed as linear combinations of the eigenfunctions $|\phi_{nm})$:
\begin{equation}
\label{eq: 20}
|P_k)=\sum_{n=0}^\infty \sum_{m=0}^\infty p^{(k)}_{nm}\,|\phi_{nm})
\ .
\end{equation}
In general these eigenfunctions are neither normalised, nor do they form an orthogonal set, but these properties are not required in the following arguments. We first consider the implications of the requirement that $\langle x_1\rangle=0$. Using (\ref{eq: 18}) and (\ref{eq: 19}), by an inductive argument involving repeated integration by parts we have:
\begin{equation}
\label{eq: 21}
\int_{-\infty}^\infty {\rm d}x_1\int_{-\infty}^\infty {\rm d}x_2\ \phi_{nm}(x_1,x_2)\,x_1
=\delta_{n1}\delta_{m0}
\end{equation}
so that the condition $\langle x_1\rangle=0$ is satisfied by requiring that $p^{(k)}_{10}=0$ in (\ref{eq: 20}) for all $k$.

\begin{figure}
\centerline{\includegraphics[width=5.0cm]{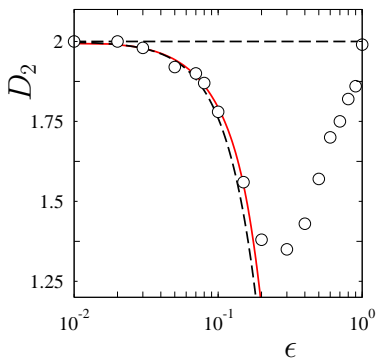}}
\caption{\label{fig: 1} Correlation dimension $D_2$ of the model (\ref{eq: 2}), as a function of the dimensionless measure of inertia $\epsilon$, defined by (\ref{eq: 13}). Here $\Gamma=3$ (incompressible flow) and $\tau\to 0$ (rapidly fluctuating flow field). The numerical data ($\circ$) are compared to the quadratic approximation of (\ref{eq: 23}), (dotted line) and to the Borel summation of the series (\ref{eq: 22}), (solid line, red online). The summation was evaluated using the method in \cite{LeG+80}, using the conformal map $z=2u^\nu/s(1-u)^\nu$, with $\nu=1/4$, $s=25$. The results converge as the number of terms used in the Borel summation, $k_{\rm max}$, increases: the curves for $k_{\rm max}=10$, $20$, $30$, $40$, $50$ lie on top of each other.}
\end{figure}

Substituting (\ref{eq: 16}) into (\ref{eq: 15}) leads to a recursion giving $|P_n)$ in terms of all of the preceding approximations: the term of order $\epsilon^n$ is
\begin{eqnarray}
\label{eq: 22}
\!\!\!\!\! 0&\!=\!&\hat F_0 |P_n)+[\hat G-\alpha_0(\hat a_1+\hat b_1)]\,|P_{n-1})\ldots
\nonumber \\
\!\!\!\!\! &\!-\!&\alpha_j(\hat a_1+\hat b_1)|P_{n-1-j})\ldots-\alpha_{n-1}(\hat a_1+\hat b_1)|P_0)
\ .
\end{eqnarray}
There are two unknowns in this equation, $|P_n)$ and $\alpha_{n-1}$; all of the other $|P_j)$ and $\alpha_j$ are assumed to have been determined at previous iterations. For any value of $\alpha_{n-1}$, equations (\ref{eq: 22}) can be solved formally for $|P_n)$ by multiplying by $\hat F_0^{-1}$. For a state $|Q)$ with coefficients $q_{nm}$ we have $\hat F_0^{-1}|Q)=-\sum_{n=0}^\infty \sum_{m=0}^\infty \frac{1}{n+m}q_{nm}|\phi_{nm})$. The action of $\hat F_0^{-1}$ upon a general state $|Q)$ is therefore undefined unless the coefficient $q_{00}$ is equal to zero. At each order we can solve (\ref{eq: 22}) for $|P_n)$ choosing the value of $\alpha_{n-1}$ so that $p^{(n)}_{10}=0$. Note that the operator $\hat G$ contains raising operators as left factors, so that $\hat F_0^{-1}\hat G|f)$ exists for any state $|f)$. However, because there is a lowering operator $\hat b_1$ acting on the states $|P_k)$, the action of multiplying the terms in (\ref{eq: 22}) by $\hat F_0^{-1}$ is only defined if all of the $|P_k)$ are chosen so that $p^{(k)}_{10}=0$. However, we have already seen that this is precisely the condition to ensure that the solution satisfies $\langle x_1\rangle=0$, that is, the solvability condition upon (\ref{eq: 22}) coincides with the condition (\ref{eq: 13}). The generation of the series (\ref{eq: 16}) was automated using an algebraic manipulation program. Iterating the equation (\ref{eq: 22}) using the initial condition $|P_0)=|\phi_{00})$ leads to the following series expansion for $D_2(\epsilon)$:
\begin{equation}
\label{eq: 23}
D_2=\Gamma-1-\Gamma(\Gamma^2-1)\epsilon^2+\Gamma(\Gamma^2-1)(3\Gamma^2+2\Gamma-11)\epsilon^4
+O(\epsilon^6)
\ .
\end{equation}
All $\alpha_j$ with odd $j$ are equal to zero, and all the coefficients are zero when $\Gamma=1$. For $\Gamma=3$ (so that $\mbox{\boldmath$\nabla$}\cdot\mbox{\boldmath$u$}=0$) the first few non-vanishing coefficients are $2$, $-24$, $528$, $-28800$, $1654848$, $-128860416$, so that the series is clearly divergent with alternating signs.
It is interesting to consider whether this series contains a complete description of $D_2(\epsilon)$. We investigated its evaluation by means of a Borel summation technique described in \cite{LeG+80}. The Borel transform $B(z)=\sum_{k=0}^\infty (\alpha_k/k!)z^k$ of $D_2(\epsilon)$ is convergent inside a disc (of radius $1/12$), but inversion of $B(z)$ to yield $D_2(\epsilon)$ requires its Laplace transform, which is an integral over $z\in (0,\infty)$. This is facilitated by making a conformal transformation to a new variable $u$, defined by $z=2^\nu u/s(1-u)^\nu$ (where $\nu$, $s$ are constants), so that the positive $z$ axis is mapped to the interval $u\in (0,1)$. We find that the expansion of $B(z)$ as a series in $u$ has decreasing coefficients when $\nu=\frac{1}{4}$ and $s=25$ (indicating that $B(z)$ is analytic in the image of the disc $|u|<1$). Performing the integral in the $u$ variable gives a summation of the series which converged as the number of terms, $k_{\rm max}$, was increased. Figure \ref{fig: 1} illustrates the results for $\Gamma=3$. For small $\epsilon$ there is excellent convergence to a numerical evaluation of $D_2(\epsilon)$. For large $\epsilon $, however, while the Borel summation converges as $k_{\rm max}$ is increased, it diverges from the numerical evaluation. This indicates that there is a component of $D_2(\epsilon)$ which has no representation as an analytic function. Non-perturbative approaches to equation (\ref{eq: 12}) are required to describe this non-analytic contribution.

Equation (\ref{eq: 7}) can be used to determine the correlation dimension of other stochastic dynamical systems, including cases where the random component has a finite correlation time, and also for deterministic systems. A full account of the use of equation (\ref{eq: 7}) to determine the correlation dimension will be published elsewhere.

{\em Acknowledgments.} This work was supported by the project \lq Nanoparticles in an interactive environment' at G\"oteborg university, and BM was supported by the Vetenskapsr\aa{}det.


\begin{thebibliography}{}

\bibitem[Shaw (2003)]{Sha03}
R. A. Shaw,
{\it Ann. Rev. Fluid Mech.}, {\bf 35}, 183-227, (2003).
%
\bibitem[Beckwith, Henning \& Nakagawa(2000)]{Beck+00}
S, V. W. Beckwith, T. Henning  and Y. Nakagawa, in {\sl Protostars and
Planets IV}, eds. V. Manning, A. P. Boss, and S. Russell (Tucson: Univ. Arizona
Press), 533, (2000).
%
\bibitem[Wilkinson, Mehlig \& Uski, (2008)]{Wil+08}
M. Wilkinson, B. Mehlig and V. Uski,
{\it Astrophys. J. Suppl.}, {\bf 176}, 484-96, (2008).
%
\bibitem[Maxey (1987)]{Max87}
M. R. Maxey,
{\it J. Fluid Mech.}, {\bf 174}, 441-65, (1987).
%
\bibitem[Ott (2002)]{Ott02}
E. Ott, {\sl Chaos in Dynamical Systems}, 2nd edition, Cambridge: University Press, (2002).
%
\bibitem[Sommerer \& Ott(1993)]{Som+93}
J. C. Sommerer and E. Ott,
{\it Science}, {\bf 259}, 335-9, (1993).
%
\bibitem[Falkovich, Fouxon and Stepanov (2002)]{Fal+02}
G. Falkovich, A. Fouxon and M. G. Stepanov,
{\it Nature}, {\bf 419}, 151-4, (2002).
%
\bibitem[Wilkinson \& Mehlig (2005)]{Wil+05}
M. Wilkinson and B. Mehlig,
{\it Europhys. Lett.}, {\bf 71}, 186-92, (2005).
%
\bibitem[Wilkinson {\em et al}(2007)]{Wil+07}
M. Wilkinson, B. Mehlig, S. \"Ostlund and K. P. Duncan,
{\it Phys. Fluids}, {\bf 19}, 113303, (2007).
%
\bibitem[Bec {\em et al}(2006)]{Bec+06}
J. Bec, L. Biferale, G. Boffetta, M. Cencini, S. Musachchio, and F. Toschi,
{\it Phys. Fluids}, {\bf 18}, 091702, (2006).
%
\bibitem[Kaplan \& Yorke(1979)]{Kap+79}
J. L. Kaplan and J. A. Yorke, in {\sl Functional Differential Equations and
Approximations of Fixed Points}, Lecture Notes in Mathematics, eds.
H.-O. Peitgen and H.-O. Walter, Springer, Berlin, {\bf 730}, 204, (1979).
%
\bibitem[Grassberger \& Procaccia (1984)]{Gra+84}
P. Grassberger and I. Procaccia,
{\it Physica D}, {\bf 13}, 34-54, (1984).
%
\bibitem[Bec, Gaw\c edzki \& Horvai(2004)]{Bec+04}
J. Bec, K. Gaw\c edzki and P. Horvai,
{\it Phys. Rev. Lett.}, {\bf 92}, 224501, (2004).
%
\bibitem[Bec (2007)]{Bec+07}
J. Bec, M. Cencini and R. Hillerbrand,
{\it Physica D}, {\bf 226}, 11-22, (2007).
%
\bibitem[Bec (2007)]{Bec+07a}
J. Bec, L. Biferale, M. Cencini, A. Lanotte, S. Musacchio and F. Toschi,
{\it Phys. Rev. Lett.}, {\bf 98}, 084502, (2007).
%
\bibitem[Mehlig \& Wilkinson (2004)]{Meh+04}
B. Mehlig and M. Wilkinson,
{\it Phys. Rev. Lett.}, {\bf  92}, 250602, (2004).
%
\bibitem[LeGuillou \& Zinn-Justin (1980)]{LeG+80}
J. C. LeGuillou and J. Zinn-Justin,
{\it Phys. Rev. B}, {\bf 21}, 3976-98, (1980).
%
\end{thebibliography}
\end{document}